\documentclass[a4paper,11pt]{article}
\pdfoutput=1 
\usepackage{amssymb,amsmath}
\usepackage{jcappub} 
\usepackage{lineno}
\usepackage[T1]{fontenc} 

\title{\boldmath Measurement of cosmic muon-induced events in an HPGe detector using time-coincidence technique}

\author[1]{Roni~Dey,\note{Corresponding author: Roni Dey, neuphyroni@gmail.com}}
\author{Dipanwita~Mondal,}
\author{Sudipta~Das,}
\author{Varchaswi~K.~S.~Kashyap,}
\author{and Bedangadas~Mohanty}

\affiliation{National Institute of Science Education and Research, An OCC of Homi Bhabha National Institute, Jatni 752050, India}
\emailAdd{neuphyroni@gmail.com}
\emailAdd{vkashyap@niser.ac.in}

\abstract{Detailed understanding and suppression of backgrounds are among the key challenges faced by Coherent Elastic Neutrino-Nucleus Scattering (CE\ensuremath{\nu}NS) experiments. The sensitivity of these experiments is largely determined by the background levels arising from various sources. Above-ground and shallow-overburden neutrino experiments typically employ passive shielding, primarily composed of lead (Pb), to suppress environmental $\gamma$ background. However, such shielding can introduce additional backgrounds that are particularly challenging for CE\ensuremath{\nu}NS experiments. These backgrounds arise mainly from $\gamma$ and neutrons produced by cosmic muon interactions in the shielding, and their contribution can become significant depending on the amount of Pb shielding used. In the current work, we measure the yield of secondary particles originating from Pb as a result of high-energy cosmic muon interaction, using a high-purity germanium (HPGe) detector and plastic scintillators. A time-coincidence technique is used to identify and reject these secondary background events from the experimental data. The obtained mean characteristic time of these residual background events is 11 $\pm$ 4 $\mu$s, which is consistent with the Geant4-based MC simulation result of 11 $\pm$ 1 $\mu$s. The measured efficiency-corrected rate of muon-induced events in the HPGe detector is 34 $\pm$ 1 (stat.) $\pm$ 3 (sys.) day$^{-1}$kg$^{-1}$ within the energy range of 30 keV to 2000 keV. The yield of muon-induced secondary backgrounds in 10 cm thick Pb shielding is evaluated to be $(11 \pm 1 (\text{ stat.}) \pm 1 (\text{ sys.}))$ secondary events$\thinspace\text{kg}^{-1}\thinspace\mathrm{m^{-2}}\thinspace\text{muon}^{-1}$ at sea level.}

\begin{document}
\maketitle
\flushbottom

\section{Introduction}
\label{sec1}
\begin{figure}
  \centering
  \includegraphics[width=0.35\textwidth]{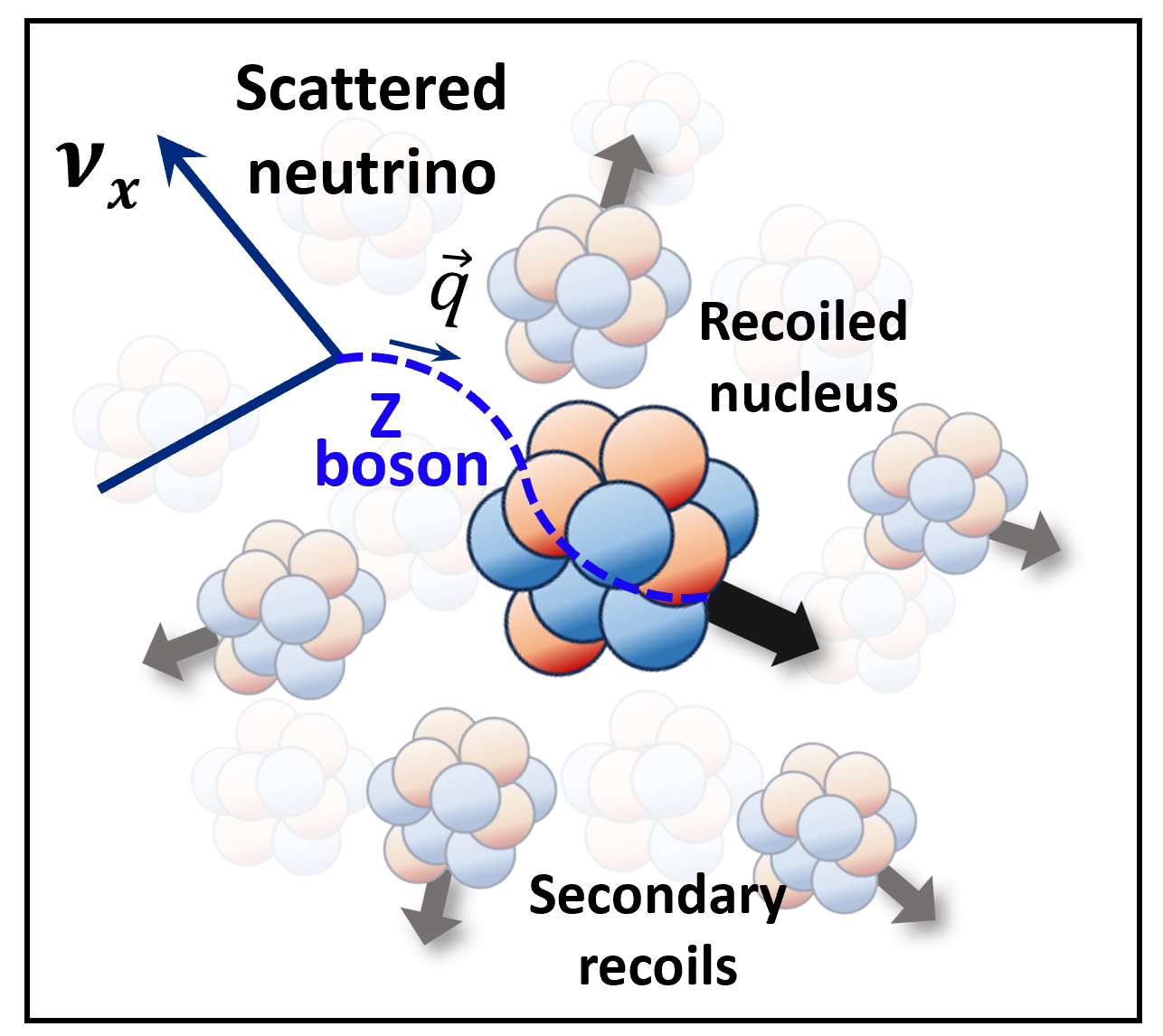}
  \caption{\label{fig1}An illustration of a neutrino scattering with a nucleus via CE\ensuremath{\nu}NS process.}
\end{figure}

Coherent Elastic Neutrino-Nucleus Scattering (CE\ensuremath{\nu}NS) is a fundamental weak neutral current interaction in standard model (SM), where a neutrino coherently scatters off a target nucleus $A(Z,N)$ via the exchange of a $Z^{0}$ boson (see equation~\eqref{CEvNS}~\cite{cevns}). This interaction results in a nuclear recoil with an energy range up to $\mathcal{O}(\text{keV})$, as illustrated in Fig.~\ref{fig1}. At low momentum transfer, the cross-section of this process scales with the square of the neutron number ($N^{2}$) of the target nucleus. The CE\ensuremath{\nu}NS interaction was experimentally observed for the first time in 2017 by the COHERENT Collaboration~\cite{coherent} using a Spallation Neutron Source (SNS). The CE\ensuremath{\nu}NS process exhibits a significantly larger cross-section due to quantum-mechanical coherency effects, compared to inverse $\beta$ decay (IBD) and neutrino-electron scattering interactions at the same energy range~\cite{coherency}. However, detecting CE\ensuremath{\nu}NS is inherently challenging due to the extremely low-energy nuclear recoil signal, which is often obscured by a substantial background event rate~\cite{conus+}. Hence, proper understanding and effective mitigation of residual backgrounds becomes important to improve the sensitivity of these experiments. CE\ensuremath{\nu}NS experiments are typically done on the surface and near nuclear reactor facilities~\cite{conus,texono}. The achievement of ultra-low background levels under such conditions presents a significant challenge. This limitation, can be mitigated by deploying a compact shell-like shielding configuration designed to effectively suppress the relevant external background sources, such as environmental $\gamma$ and reactor-correlated backgrounds~\cite{texono_bkg}. 


\begin{equation}\label{CEvNS}
  \nu + A(Z,N) \rightarrow \nu + A(Z,N)     
\end{equation}

A major background arises due to cosmic muon-induced secondary $\gamma$ and neutrons within the compact high-Z shielding structure, which could significantly limit the sensitivity for rare event search experiments. High energy cosmic muons can generate electromagnetic cascades and neutrons as they pass through high-Z materials, such as lead (Pb)~\cite{mu_lead1,mu_lead2}. Moreover, in deep underground facilities, muon-induced secondary particles originating from the rock overburden contribute to the background~\cite{wimp,dm_bkg}. Since these backgrounds cannot be actively rejected, they can easily mimic CE$\nu$NS interactions. Therefore, detailed studies are performed to quantify the rate of these secondary backgrounds using typical shielding configurations in sec.~\ref{sec2}, employed in rare event search experiments. Suppression of muon-induced secondaries in experimental data can be achieved through the use of different configurations of passive shielding as well as an active muon veto system. 


The radioanalytical laboratory at NISER is equipped with a commercial coaxial p-type HPGe detector with a passive $\gamma$ shield consisting of Pb and copper (Cu). Two plastic scintillator detectors, were arranged to make a telescope on top of this setup for triggering cosmic muons. Using this arrangement, the muon-induced secondary $\gamma$ and neutron production rates has been measured. A time coincidence technique has been used to select and mitigate the muon induced events. Simulations have also been performed using Geant4 to reproduce the experimental results. 

The article is organized as follows: Section~\ref{sec2} provides a detailed description of the experimental setup and the data acquisition system. The time correlation of cosmic muons and muon-induced secondary particles, as obtained from a Geant4-based MC simulation, is discussed in detail in Section~\ref{sec3}. The characterization of the HPGe detector are presented in Section~\ref{sec4}. This section also includes detector response modeling and a comparison of the results with simulations. In Section~\ref{sec5}, we report measurements of cosmic muons and muon-induced secondary events in the Pb shield surrounding the HPGe detector, and explore the possibility of discriminating these events based on their timing characteristics. The evaluated production yields of the secondary $\gamma$ and neutrons are discussed in detail.

\section{Experimental Setup and Data Acquisition System}
\label{sec2}
\begin{figure}
  \centering
  \includegraphics[width=8.4cm,height=4.8cm]{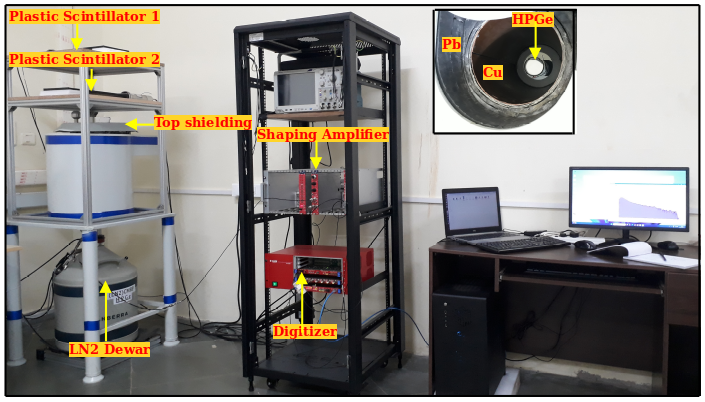}
  \caption{\label{fig2}Experimental setup and DAQ system at the low-background laboratory at NISER.}
\end{figure}

The experimental setup consists of a coaxial p-type HPGe detector, CANBERRA make (Model: GC3018), weighing $\sim$1 kg having a crystal diameter of 61.7 mm and a height of 40 mm, specially designed for high resolution gamma spectroscopy~\cite{canberra}. The detector is housed in a vacuum cryostat  coupled to a liquid nitrogen dewar, and is enclosed within a 1.5 mm thick aluminum (Al) casing. The passive shielding comprises a cylindrical 10 cm thick Pb layer, lined on the inside with 1.5 mm thick cadmium, 2 mm thick Al and 1.5 mm thick Cu. The top of the shield has a 10 cm thick sliding Pb door to enable sample placement and removal within the shield. The inner diameter of the shield is 300 mm and outer diameter is 500 mm. The total mass of the passive shielding is $\sim$600 kg. Two plastic scintillator detectors (PSs), each measuring 30 cm $\times$ 18 cm $\times$ 1 cm, are installed on top of the movable shield, positioned 30 cm apart, serving as a compact muon telescope veto system. The shielding and scintillator telescope is mounted on a stainless steel (SS) table. The HPGe detector and cryostat assembly is placed below the table, coaxial to the shielding as shown in Fig.~\ref{fig2}. The placement is such that the HPGe is surrounded by the shielding from all directions, except the bottom side. A reverse bias voltage of $2.9$ kV was applied to the HPGe detector. The HPGe detector, enclosed within the shielding, measured an integral ambient background count rate of $\sim$ $\mathrm{10^{6}}$ day$^{-1}$kg$^{-1}$ at sea level, in the energy range of 30 keV to 2000 keV.

The HPGe detector has an integrated charge-sensitive preamplifier (Model: IPA-SL10). The signal from the preamplifier is taken to the data acquisition system that includes a NIM-based high voltage power supply, spectroscopic amplifier (Model: N968), and a digitizer ~\cite{caen}. The data acquired from the HPGe detector, is recorded on an event-by-event basis with a timestamp and a energy deposition. A spectroscopic amplifier has been used to achieve a fast time response with a shaping time of 0.5 $\mu$s and a gain of 20, converting the long exponential decay tail of the pulse into a Gaussian-like pulse. A commercial CAEN VME-based V1730 digitizer is used to record waveform with 14-bit resolution at a sampling rate of 500 MS$\thinspace$s$^{-1}$. The signals from all three detectors (HPGe + 2 scintillators) are independently acquired and processed using the CAEN charge integration firmware and the CoMPASS software~\cite{compass}. The time coincidence between the events in the HPGe detector and muon telescope veto system are analyzed using offline algorithms developed in ROOT and C++~\cite{root}.

\section{\label{sec3}Geant4-based MC simulation for studying muon-induced secondaries}
A Geant4-based MC simulation framework has been developed to model the experimental setup, including the detector geometry, detector material, muon telescope veto system, and compact layered shielding structure. The simulation incorporates various physics processes in PhysicsList for different particle interactions, such as muon-nuclear, electromagnetic, and hadronic interactions. This comprehensive MC simulation study aims to benchmark our experimental results and estimate the event reconstruction efficiency.

\subsection{\label{sec3.1}Description of PhysicsList and CRY Event Generator}
The full experimental geometry was modelled using the Geant4 toolkit (version 11.1.2)~\cite{geant4} as shown in Fig.~\ref{fig3}(a). The Geant4 simulation package offers a variety of physics lists, which are pre-defined sets of physics interactions that can be customized to achieve specific simulation requirements. This study primarily requires standard electromagnetic and radioactive decay physics processes to obtain the energy response model of $\gamma$, electrons and positrons from radioactive sources in the HPGe detector. Additionally, the high precision \texttt{QGSP\_BIC\_HP} physics model is incorporated in the user-defined physics list to describe high-energy hadron-nucleus and hadron-nucleon interactions. 
In order to accurately estimate the production rate of secondary particles, the physics list explicitly incorporates Photo-Nuclear interaction (\texttt{G4PhotoNuclearProcess}), Muon Capture process (\texttt{G4MuonMinusCapture}), Muon-Nucleus interaction (\texttt{G4MuonNuclearProcess}), and Muon Decay physics processes.

High-energy cosmic muons can be simulated using various open-source generators, which can be integrated into particle transport frameworks like Geant4. In the present work, cosmic muons ($\mu^+$ and $\mu^-$) are generated at sea level using the Cosmic-Ray Air Shower (CRAS) generator, with energies ranging from 100 MeV to 100 GeV and an angular distribution following $\cos^2\theta$, with a charge ratio of $\mu^+/\mu^- = 1.3$. This generator contains the Cosmic Ray Shower library (CRY) and is integrated with Geant4 as the particle transport framework~\cite{cry}. 

\subsection{\label{sec3.2}Expected backgrounds for above-ground CE$\nu$NS experiments}


The dominant backgrounds in reactor-based and above-ground CE\ensuremath{\nu}NS experiment can be classified into three categories: 
\begin{itemize}
  \setlength{\parskip}{0pt}
  \setlength{\itemsep}{0pt}
  \item[(a)] Environmental background,
  \item[(b)] Reactor-induced background,
  \item[(c)] Cosmogenic background.
\end{itemize}

The primary sources of environmental background are $\gamma$-rays, predominantly from $\mathrm{{}^{40}K}$ at 1460.8 keV and $\mathrm{{}^{208}Tl}$ at 2614.5 keV, which dominate the energy spectrum up to 3000 keV. The environmental $\gamma$-rays and reactor generated neutrons and $\gamma$ backgrounds can be mitigated by the use of layered passive shielding structure. However, the most challenging backgrounds to discriminate are the cosmogenic backgrounds. In the present study, we focus on the measurement of these cosmogenic backgrounds and discuss mitigation strategies based on time-coincidence techniques. A high energy cosmic muon can generate secondary $\gamma$/neutrons, while traversing through high-$Z$ shielding materials via multiple processes. Some of these processes are described below.


\begin{enumerate}
  
  \item \textbf{Bremsstrahlung $\gamma$:} A high energy cosmic muon interacts with Pb and generates bremsstrahlung $\gamma$ due to electromagnetic interactions. Moreover, if the muons decay inside the shielding, the resulting $e^-/e^+$ from the decay can also generate Bremsstrahlung $\gamma$~\cite{brem_bkg}.
  
  \item \textbf{Photo-nuclear:} In photo-nuclear interactions, a high-energy muon interacts with a nucleus via the exchange of a virtual photon (electromagnetic interaction), leading to nuclear disintegration or the emission of hadrons. Photoproduction primarily occurs within electromagnetic cascades induced by cosmic muons through bremsstrahlung and $e^+e^-$ pair production, resulting to the generation of neutrons. This mechanism serves as a primary channel for muon-induced neutron backgrounds in deep underground laboratories conducting rare event search experiments~\cite{pho_bkg}. Additionally, hadroproduction primarily originates from the hadronic cascades caused by muon inelastic scattering, also known as muon-nuclear interaction (Deep inelastic scattering).
  
  \item \textbf{Muon spallation:} As high energy cosmic muons ($\sim$GeV) traverse matter, their energy loss can lead to nuclear breakup processes, referred to as ``spallation''. These interactions can produce relatively long-lived radioactive isotopes directly through the spallation of stable nuclei. Additionally, muon interactions can generate major secondary particles as spallation products, such as gamma, electrons, neutrons, pions, and other mesons~\cite{spa_bkg}. 
  
  \item \textbf{Muon capture:} Low energy negative muons ($\sim$ 100 MeV) undergo Coulomb interaction, forming a ``muonic atom'' bound state, resulting in the production of a neutrino and a neutron ($\mathrm{\mu^{-}}$ + p $\mathrm{\rightarrow}$ n + ${\ensuremath{\nu}}_{\mu}$). Heavy nuclei, such as Pb, have lower neutron separation energies compared to lighter nuclei, resulting in a higher neutron multiplicity per capture event~\cite{cap_bkg_1}. 
  
  
  
  
\end{enumerate}

Furthermore, for reactor-based experiments, neutron emission can arise from $(\alpha,n)$ reactions occurring in the concrete walls and environment of the reactor building, as well as from spontaneous fission of heavy isotopes~\cite{rec_bkg}.



\subsection{\label{sec3.3}Simulation of cosmic muon and muon-induced backgrounds}
\begin{figure}
  \centering
  \includegraphics[width=11.2cm,height=6.0cm]{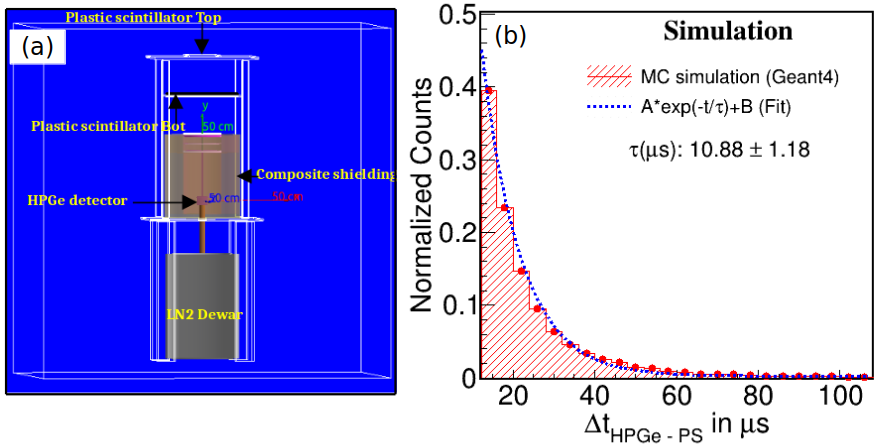}
  \caption{\label{fig3}(a) Geant4-based simulation of the HPGe detector, composite shielding, and muon telescope system. (b) Timestamp difference distribution ($\mathrm{\Delta t_{HPGe-PS}}$) between muon veto system and HPGe detector in Geant4 MC simulation. The distribution is fitted with an exponential function.}             
\end{figure}

A total of 3 million muons were randomly generated on a plane above the muon telescope veto system using the CRY software. CRY simulates realistic momentum and angular distributions of cosmic muons at sea level~\cite{mu_lead2}. These muons then pass through the muon veto system followed by the HPGe detector, which is surrounded by the shielding setup. Muon interactions with Pb generate secondary $\gamma$ and neutrons, through various physics processes, as described in ~\ref{sec3.2}. The subsequent passage of these secondary particles through the HPGe detector is also modeled using Geant4. Figure~\ref{fig3}(b) shows the distribution of the difference in timestamps ($\Delta t_\mathrm{HPGe-PS}$) between events recorded in the muon veto system and the HPGe detector from simulation. The $\Delta t_\mathrm{HPGe-PS}$ distribution decreases exponentially, due to the contribution of the secondary events, becoming negligible beyond $\sim$100 $\mu$s. It is fitted with a combined function consisting of an exponential term for muon-induced events and a constant term. The mean characteristic time ($\tau_\mathrm{sim}$) obtained from the fit is 11 $\pm$ 1 $\mathrm{\mu s}$.

\section{\label{sec4}Evaluation of spectroscopic performance of HPGe detector}
\begin{figure}
  \centering
  \includegraphics[width=15.6cm,height=5.0cm]{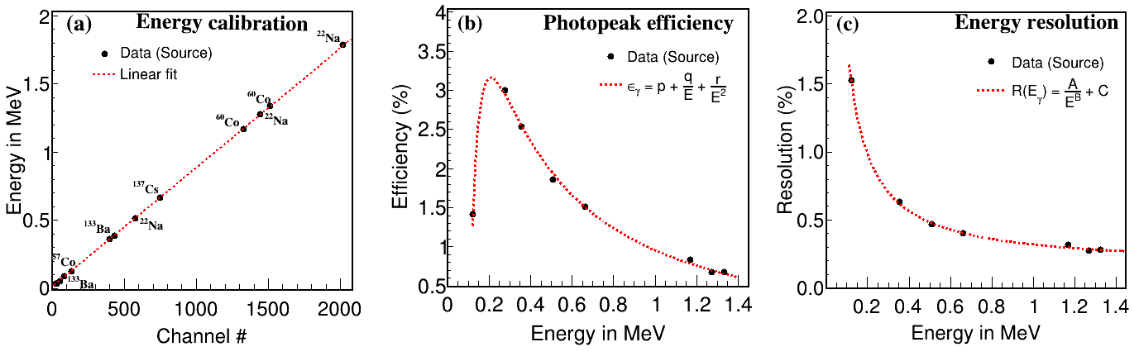}
  \caption{\label{fig4}(a) Energy calibration of the HPGe detector using standard radioactive sources. (b) Efficiency curve of the HPGe detector at source to detector distance of z = 3 cm. (c) Energy resolution model of the HPGe detector.}      
\end{figure}

The accuracy of the measurements is governed by the performance and stability of the HPGe detector and its associated electronics. Different characteristics of the HPGe detector, such as energy calibration, energy resolution, and photopeak efficiency are determined as a function of incident $\gamma$ energy. The HPGe detector has been calibrated using different standard radioactive sources such as $^{133}$Ba, $^{137}$Cs, $^{57}$Co, $^{60}$Co, and $^{22}$Na as shown in Fig.~\ref{fig4} (a). The data points are fitted with a linear function ($y = a + bx$), showing a good linear correlation between the channel number and the known energies of the calibration sources in the range of 30 keV to 2000 keV, where a = 0.00562 and b = 0.000875, with a $\mathrm{\chi^2/NDF}$ = 0.95. Below 30 keV, we observed a non-linear response in the HPGe detector due to $\gamma$-ray attenuation in the Al casing surrounding the germanium crystal. 




The absolute photopeak efficiency is defined as the ratio of the number of counts in a photopeak to the number of $\gamma$ emitted by the source. Since efficiency varies with the $\gamma$-ray energy ($E_{\gamma}$), it can be determined using the formula,
\begin{equation}\label{eq:efficiency1}
  \epsilon_{\gamma} = \frac{N_{\gamma}}{I_{\gamma} \times A(t) \times \Delta t}\ , 
\end{equation}
where $I_{\gamma}$ is the $\gamma$ emission probability (absolute intensity), $N_{\gamma}$ is the number of counts under the photopeak area after ambient background subtraction, $A(t)$ is the activity of source at the time of performing the experiment and $\Delta t$ is the data collection time interval. Figure ~\ref{fig4}(b) presents the energy-dependent efficiency of the detector at a vertical source-to-detector distance of $z = 3$ cm. The empirical formula used to fit the efficiency as a function of photopeak energy $E_{\gamma}$, is adopted from reference~\cite{Eff},
\begin{equation}\label{eq:efficiency2}
  \epsilon_{\gamma}(E_{\gamma}) = p + \frac{q}{E_{\gamma}} + \frac{r}{E_{\gamma}^{2}}\ ,
\end{equation}
where $\mathrm{\epsilon_{\gamma}(E_{\gamma})}$ represents the absolute photopeak efficiency. The best-fit parameters are found to be $p =  -0.3524,~ q = 1.4601$ and $r = -0.1516$. At very low energies below 100 keV, the detector efficiency decreases significantly. However, the efficiency curve peaks at 180 keV, followed by a sharp decline due to the increase in incident $\gamma$ energy. This behavior indicates that the efficiency peaks at low energies due to the dominance of the photoelectric effect in the Ge crystal and decreases rapidly at higher $\gamma$-rays energies, as other physics processes, such as Compton scattering and pair production, become more significant. The obtained photopeak efficiency of 3$\%$ for the HPGe detector at 274 keV and $z = 3$ cm is consistent with the values reported in the literature~\cite{Eff}.

The HPGe detector exhibits excellent energy resolution, which is quantified by the Full Width at Half Maxima (FWHM) of a single energy peak at a specific energy, where, FWHM is typically expressed in keV. HPGe detectors are widely preferred for high-resolution $\gamma$ spectroscopy, despite their relatively low efficiency at lower energies compared to inorganic scintillation detectors like NaI~\cite{Eff}. Furthermore, the very ﬁne energy resolution of HPGe detectors significantly reduces systematic uncertainties in photopeak analysis, especially in the presence of background continua. This is particularly important in scenarios where muon-induced $\gamma$ spectra are encountered. Generally, the energy resolution of an HPGe detector is expressed as the ratio of $\sigma$ ( FWHM = 2.35 $\times$ $\sigma$) to the $\gamma$ energy (Resolution = $\sigma/E_{\gamma}$). The energy resolution as a function of $\gamma$ energy from various sources is shown in Fig.~\ref{fig4}(c) and is fitted using an empirical three-parameter function~\cite{Eff}.

\begin{equation}\label{eq:efficiency2}
  \frac{\sigma}{E_{\gamma}} = \frac{A}{E_{\gamma}^{B}} + C
\end{equation}
The energy resolution of the detector was measured to be 0.4$\%$ at a $\gamma$-ray energy of 662 keV. The best-fit values of the parameters $A, B$, and $C$ are determined to be 0.0673, 1.0194, 0.0688 respectively. We added a spectroscopic amplifier with a shaping time of 0.5 $\mu$s in the DAQ chain to have better timing characteristics for the HPGe detector. However, this induced a contribution of high-frequency electronic noise to the integrated signal. As a result, the FWHM of the HPGe detector in the current study is $\sim$6 keV at a $\gamma$-ray energy of 662 keV, which is slightly worse than the typically expected value of $\sim$2 keV at 662 keV energy~\cite{reso}. 

\begin{figure}
  \centering
  \includegraphics[width=6.2cm,height=5.4cm]{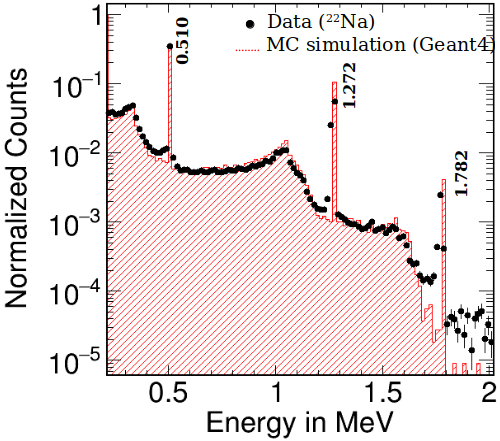}
  \caption{\label{fig5}Comparison between the measured and simulated energy response of the HPGe detector using a $\mathrm{{}^{22}Na}$ source, placed at a verticle distance of z = 3 cm.}      
\end{figure}

The $\gamma$ spectrum of a $\mathrm{{}^{22}Na}$ source, placed 3 cm vertically above the detector, was measured using the HPGe detector. It was then compared with Geant4 simulation. The user-defined physics list in Geant4 incorporated both the \texttt{G4RadioactiveDecayPhysics} and \texttt{G4EmStandardPhysics} processes. Additionally, the measured energy-dependent resolution of the HPGe detector was applied to the simulated energy spectrum. Finally, the simulated spectrum was compared with the measured response after the background subtraction, as shown in Fig.~\ref{fig5}. A good agreement between data and simulation was observed across the entire energy range including the Compton edge~\cite{comp}. This comprehensive Geant4-based simulation study serves as a valuable benchmark for experimental investigations of muon-induced secondary background in HPGe detector.


\section{\label{sec5}Measurement of muon-induced secondary $\gamma$ and neutrons}

\subsection{\label{sec5.1}Detection methodology}
\begin{figure}
  \centering 
  \includegraphics[width=7.0cm,height=6.4cm]{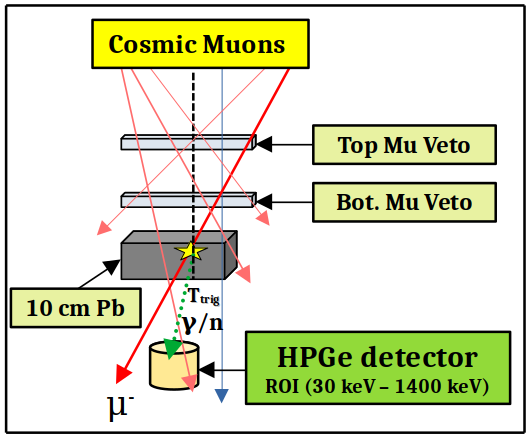}
  \caption{\label{fig6}Illustration of the detection technique for muon-induced secondaries.}      
\end{figure}

The muon telescope veto system consists of two plastic scintillator paddles of dimension $\mathrm{30\times 18\times 1}$ $\mathrm{cm^{3}}$. The paddles are arranged in a staggered configuration, 30 cm apart from each other, covering a surface area of 324 $\mathrm{cm^{2}}$, and are used to trigger cosmic muon events. A schematic diagram of the detection technique of cosmic muon-induced secondries is illustrated in Fig.~\ref{fig6} and described as follows:

\begin{enumerate}
  \item Events traversing both plastic scintillator paddles in the muon telescope veto system are identified as triggered cosmic muons. \par
  
  \item Triggered cosmic muons may interact with the Pb in the movable top shielding, generating $\gamma$ and neutrons. 
  
  \item These secondary particles can deposit energy in the HPGe detector. The HPGe detector measures energy depositions in the range of 30 keV to 2000 keV. \par
  
  \item Reconstructed muon-induced secondary events are analysed within a predefined coincidence time window of 300 $\mu$s, following the passage of a cosmic muon through the muon telescope veto system. \par
  
  \item A similar analysis is carried out without the top shielding configuration to evaluate the contribution of the Pb shielding to the observed secondary events.
  
\end{enumerate}

\subsection{\label{5.2}Cosmic muon rate in the muon telescope veto system}
\begin{figure}
  \centering
  \includegraphics[width=5.8cm,height=5.2cm]{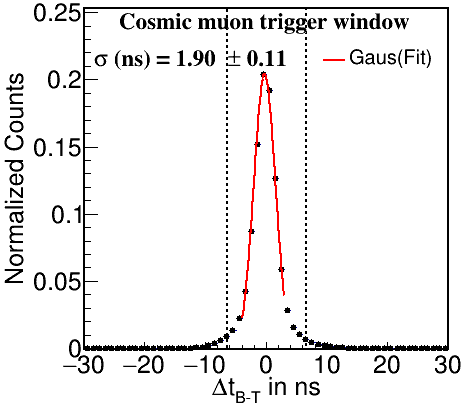}
  \caption{\label{fig7}Timestamp difference distribution ($\mathrm{\Delta t_{B-T}}$) between the two trigger paddles. The muon events are selected within the dotted region for the current analysis.}     
\end{figure}

The muon telescope system detects high-energy cosmic muons passing through both the top and bottom scintillator paddles, as shown in Fig.~\ref{fig6}. The timestamps of events from both paddles are recorded independently and denoted as $t_\mathrm{Top}$ and $t_\mathrm{Bottom}$, respectively. The distribution of the timestamp difference ($\Delta t_\mathrm{B-T}$) between $t_\mathrm{Top}$ and $t_\mathrm{Bottom}$ is presented in Fig.~\ref{fig7}. A peak between $\pm$ 6 ns in the $\Delta t_\mathrm{B-T}$ distribution is observed, indicating coincidences associated with the passage of cosmic muons through both scintillator paddles. Consequently, muon-triggered events are selected for this analysis within a coincidence time window of $\pm$ 6 ns in the $\Delta t_\mathrm{B-T}$ distribution, referred to as the cosmic muon-trigger window. This selection of the trigger window effectively reduces accidental $\gamma$ backgrounds. The average muon-triggered event rate recorded by the muon telescope veto system is $\sim4.8$ Hz. Outside this time window, where 6 $<$ $|\Delta t_\mathrm{B-T}|$ (ns) $<$ 12, the event rate is $\sim 0.2$ Hz~\cite{bkg_1}. This rate corresponds to accidental ambient $\gamma$-rays and cosmic muons traversing the scintillator paddles at a grazing angle that do not interact with the lead shield due to geometrical acceptance. The rate of these grazing muons is significantly lower than the overall muon-triggered rate of $\sim$ 4.8 Hz. 

\subsection{\label{5.3}Evaluation of muon-induced secondary event rate in the HPGe detector}
\begin{figure}
  \centering
  \includegraphics[width=10.4cm,height=9.2cm]{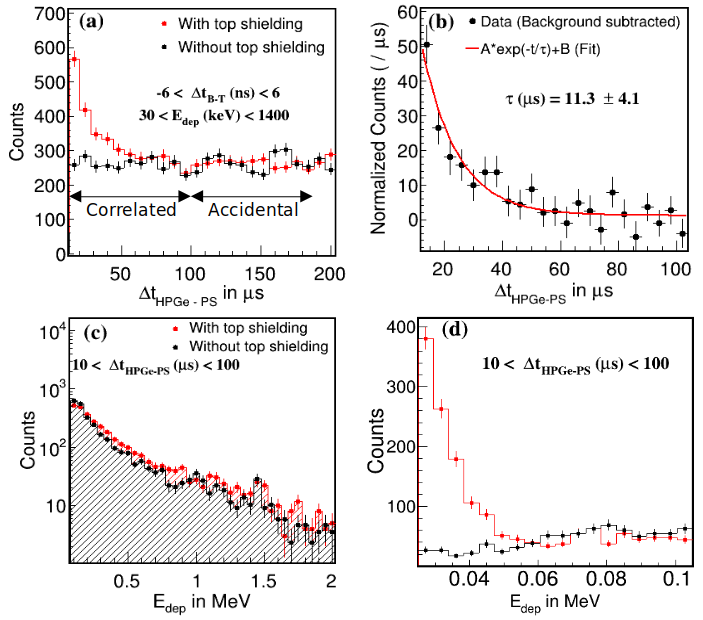}
  \caption{\label{fig8}(a) Comparison of the measured $\mathrm{\Delta t_{HPGe-PS}}$ distributions for with and without top shielding configurations. (b) The background subtracted $\mathrm{\Delta t_{HPGe-PS}}$ distribution is fitted with a combined function of an exponential and a constant. (c) Comparison of the measured energy distributions in the HPGe detector for configurations with and without top shielding within the correlated time interval, and (d) shows a zoomed-in view of the comparison in the energy range below 100 keV.}    
\end{figure}

After the selection of muon triggers, the timestamp of the top plastic scintillator paddle denoted by $t_\mathrm{Trig}$, which are referred to as prompt events. Events reconstructed in the HPGe detector with deposited energies in the range of 30 $<$ $E_\mathrm{{dep}}$ (keV) $<$ 1400 and within a $300\ {\mu}$s coincident time window are referred to as delayed events~\cite{bkg_2}. $E_\mathrm{{dep}}$ is defined as the reconstructed total energy of an event, obtained by summing the energy deposited in the Ge crystal from all interactions occurring within that time window. The upper energy threshold of 1400 keV was set to reject external $\gamma$, primarily from $\mathrm{^{40}K}$ (1460 keV) and $\mathrm{^{208}Tl}$ (2614 keV) as observed in the HPGe detector. Fig.~\ref{fig8}(a) shows the timestamp difference distributions ($\Delta t_\mathrm{HPGe-PS}$) between prompt and delayed events, comparing configurations with and without the top movable Pb shielding. These $\Delta t_\mathrm{HPGe-PS}$ distributions are shown for events within the cosmic muon-trigger window of $-6<\Delta t_\mathrm{B-T}\ \mathrm{(ns)}<+6$. The $\Delta t_\mathrm{HPGe-PS}$ distributions are divided into two time intervals (1) Correlated, ranging from 10 $<$ $\Delta t_\mathrm{HPGe-PS}$ ($\mu$s) $<$ 100, and (2) Accidental, ranging from 100 $<$ $\Delta t_\mathrm{HPGe-PS}$ ($\mu$s) $<$ 190. This selection is influenced by the spread of the $\Delta t_\mathrm{HPGe-PS}$ distributions for prompt-delay paired events from the Geant4 simulation results, which decrease to zero at $\sim$100 $\mu$s, as shown in Fig.~\ref{fig3} (b).



As shown in the $\Delta t_\mathrm{HPGe-PS}$ distributions in Fig.~\ref{fig8}(a), the number of prompt-delay paired events within the correlated time interval increases when the top Pb shielding is installed, compared to the configuration without the shielding structure. Beyond the 100 $\mu$s coincidence time window, the distributions for both configurations appear uniform and scale proportionally, indicating the randomness of the prompt-delay paired events, which corresponds to the purely accidental background. Fig.~\ref{fig8}(b) represents the $\Delta t_\mathrm{HPGe-PS}$ distribution after subtracting the `without top shielding' time spectrum from the `with top shielding' time spectrum. The background-subtracted distribution is modeled by fitting it within a time window of 10 $\mu$s to 200 $\mu$s using a combined function consisting of an exponential component, representing muon-induced secondary correlated events and a constant term accounting for random accidental background. The obtained mean characteristic time of 11 $\pm$ 4 $\mathrm{\mu}$s, which is consistent with the MC simulation result of 11 $\pm$ 1 $\mathrm{\mu}$s. A comparison of the measured energy distributions in the HPGe detector for configurations with and without top shielding, within the correlated time window, are shown in Fig.~\ref{fig8}(c). It is observed that the muon-induced secondaries predominantly deposit energy in the HPGe detector in the low-energy region below 60 keV, as shown in the Fig.~\ref{fig8}(d).



A total of 23 days of data was recorded with top shielding and 19 days without top shielding. A statistical subtraction method is employed to extract the rate of secondary $\gamma$ and neutrons produced by high-energy cosmic muons in Pb. Based on the measured data, the integrated yields of prompt-delay (PD) paired events within the correlated time interval, and the energy range of 30 keV to 1400 keV, with (W) and without (WO) top shielding, as depicted in Fig.~\ref{fig8}(a) are $N^\mathrm{W}_\mathrm{PD}\mathrm{(Corr.)}$ = 4114 $\pm$ 64 (stat.) and $N^\mathrm{WO}_\mathrm{PD}\mathrm{(Corr.)}$ = 2809 $\pm$ 53 (stat.), respectively. Similarly, the accidental yields from Fig.~\ref{fig8} (a) for the configurations with and without the shielding structures are $N^\mathrm{W}_\mathrm{PD}\mathrm{(Acc.)}$ = 3316 $\pm$ 58 (stat.) and $N^\mathrm{WO}_\mathrm{PD}\mathrm{(Acc.)}$ = 2754 $\pm$ 52 (stat.), respectively. After time normalization, the integrated yields of the prompt-delay paired events within the accidental time interval, as evaluated from $\Delta t_\mathrm{HPGe-PS}$ distributions are found to be consistent within the statistical uncertainties. 



\begin{figure}
  \centering
  \includegraphics[width=\linewidth]{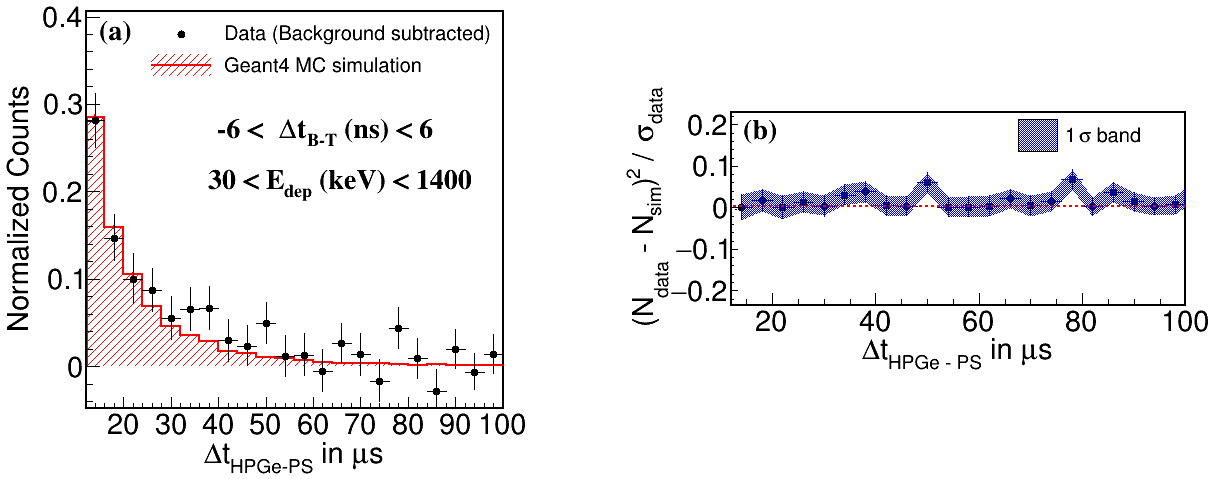}
  \caption{\label{fig9}(a) Comparison between data and Geant4-based MC simulation of the background-subtracted $\mathrm{\Delta t_{HPGe-PS}}$ distribution. (b) Bin-to-bin comparison between the measured data and MC simulation for the background-subtracted $\mathrm{\Delta t_{HPGe-PS}}$ distribution.}
\end{figure}

To validate the experimental results, a comparison with Geant4 simulation is performed. The comparison is shown in Fig.~\ref{fig9}(a). The results agree well within $1\sigma$ statistical uncertainty across the entire time range as illustrated in Fig.~\ref{fig9}(b). This consistency indicates that the MC simulation adequately reproduces the temporal behavior of muon-induced secondary events, thereby reinforcing the reliability of the experimental results and background subtraction methodology. The total number of muon-induced secondary events ($N_\mathrm{MI}$) originating from the top shield are obtained by integrating the $\Delta t_\mathrm{HPGe-PS}$ distribution of prompt-delay paired events, shown in Fig.~\ref{fig8}(b). The difference in the integrated yield of these paired events over the two time intervals (Correlated and Accidental) can be used for evaluating the correlated muon-induced secondary events, caused by the passive Pb top shielding. We follow the prescription in Ref.~\cite{bkg_sub_1, bkg_sub_2} to obtain $N_\mathrm{MI}$ using statistical subtraction method,


\begin{equation}\label{event1}
  N_\mathrm{MI}=[N^\mathrm{W}_\mathrm{PD} \mathrm{(Corr.)} -N^\mathrm{W}_\mathrm{PD}\mathrm{(Acc.)}] - K[N^\mathrm{WO}_\mathrm{PD}\mathrm{(Corr.)} -N^\mathrm{WO}_\mathrm{PD}\mathrm{(Acc.)}]\ ,
\end{equation}
where the scaling factor $K$ represents the ratio of the number of days with shielding to those without shielding, which is 1.2. The integrated yield of muon-induced secondary events is $\mathrm{N_{MI}}$ = 732 $\pm$ 27 (stat.), which correspond to 32 $\pm$ 1 events day$^{-1}$kg$^{-1}$.


\subsection{\label{5.4}Estimating the event reconstruction efficiency}

Reconstruction efficiency values were estimated based on two primary selection criteria: $\Delta t_\mathrm{HPGe-PS}$ and $E_\mathrm{{dep}}$. These criteria were applied to the experimental data to reject random $\gamma$ events with $E_\mathrm{{dep}} > 1400$ keV and $\Delta t_\mathrm{HPGe-PS} > 100$ $\mu$s. A similar cut-based analysis was performed on the simulated dataset to evaluate the corresponding reconstruction efficiency. The cumulative effect of these event selection cuts must be assessed to determine the final efficiency. A detailed efficiency estimation is summarized in Table~\ref{tab2}, outlining the selection criteria applied to the cosmic muon correlated $\gamma$ and neutron events and their respective impacts on the reconstruction efficiency. The cumulative reconstruction efficiency is determined to be $\sim93$\%. Consequently, the efficiency-corrected rate ($N^\mathrm{cor}_\mathrm{MI}$) of muon-induced secondary $\gamma$ and neutron events measured in the HPGe detector is 34 $\pm$ 1 (stat.) events per day.


\begin{table}[h]
  \small%
  \centering  
  \caption{\label{tab1}Selection cuts and their corresponding event selection efficiencies in the Geant4-based MC simulated dataset of muon-induced secondaries.\\}
  \begin{tabular}{c|c|c}
    \hline
    \bf Variables & \bf Selection criteria & \bf Efficiency ($\%$) \\ 
    \hline
    $\Delta t_\mathrm{HPGe-PS}$          &  $10<\Delta T_\mathrm{PD}\ (\mu\text{s})<100$                   &  99.1  \\
    \hline
    $E_\mathrm{{dep}}$                   &  $0.03 < E_\mathrm{{dep}}\text{ (MeV)} < 1.4$                   &  93.2  \\
    \hline
  \end{tabular}
\end{table}

\subsection{\label{5.5}Calculation of systematic uncertainties}
The following sources of systematic uncertainties were considered in the measurement of muon-induced secondary $\gamma$ and neutrons. These include uncertainties associated with (a) the reconstruction of $E_\mathrm{{dep}}$ for secondary events in the HPGe detector, (b) the choice of the $\pm$ 6 ns cosmic muon-trigger window in the muon telescope veto system, (c) the selection of coincidence timing ($\Delta t_\mathrm{HPGe-PS}$) between prompt-delay paired events, and (d) efficiency estimation. In the present study, the estimation of systematic uncertainties is carried out by varying the selection windows, referred to as set 1 and set 2, corresponding to the variables discussed above. The differences in the measured $N_\mathrm{MI}$ arising from these variations are denoted as $\sigma_{1}$ and $\sigma_{2}$, respectively. The total systematic uncertainty associated with each variable is then determined by adding these variations in quadrature as follows: $\sigma = \sqrt{(\sigma_{1}^{2} + \sigma_{2}^{2})/2}$. The selection criteria for the cosmic muon-trigger window were varied between $\pm4$ ns and $\pm8$ ns to evaluate the associated systematic uncertainty, leading to a 1$\%$ uncertainty in the measured $N_\mathrm{MI}$ events.

Similarly, the reconstruction of $E_\mathrm{{dep}}$ of delayed events is affected by the energy range chosen in the HPGe detector, between 1350 keV and 1450 keV. This variation leads to an estimated $\sim$1$\%$ change in the measured $N_\mathrm{MI}$ events. The associated uncertainty arises from variations in the reconstructed energy range, implemented to reject environmental $\gamma$ events. 

As observed from the Geant4 simulations shown in Fig.~\ref{fig3}(b), the $\Delta t_\mathrm{HPGe-PS}$ distribution follows an exponential decay, with contribution to the secondary events becoming negligible beyond $\sim$100 $\mu$s. This characteristic introduces uncertainty in the selection of the Correlated and Accidental time intervals. To estimate this uncertainty, two alternative time intervals for $\Delta t_\mathrm{HPGe-PS}$ are considered: 10 $<$ $\Delta t_\mathrm{HPGe-PS}$ ($\mu$s) $<$ 90 and 10 $<$ $\Delta t_\mathrm{HPGe-PS}$ ($\mu$s) $<$ 110. The difference in $N_\mathrm{MI}$ obtained using these intervals is used to quantify the systematic uncertainty associated with the choice of $\Delta t_\mathrm{HPGe-PS}$. The lower limit of $\Delta t_\mathrm{HPGe-PS}$ is set at 90 $\mu$s, as the Correlated-to-Accidental event ratio approaches unity at this threshold, as seen in Fig.~\ref{fig8}(a). This equivalence introduces an ambiguity in distinguishing between correlated and accidental events, thereby contributing to the uncertainty in the measured yields of secondary $\gamma$ and neutrons. Consequently, an additional systematic uncertainty of 7.1$\%$ is attributed to the selection of the $\Delta t_\mathrm{HPGe-PS}$ intervals.

A conservative systematic uncertainty of 4.5$\%$ is also assigned to the efficiency determination, based on Geant4 simulation model. Table~\ref{tab2} summarizes the systematic uncertainties associated with the selection criteria, which are propagated to the final yield of muon-induced secondaries. A total systematic uncertainty of 8.5$\%$ is estimated for the 23 days of recorded data using the HPGe detector with the shielding configuration. The final number of correlated muon-induced secondary events is obtained after correcting for the reconstruction efficiency, and is calculated to be 34 $\pm$ 1 (stat.) $\pm$ 3 (sys.) events per day in the HPGe detector, within the energy range of 30 keV to 2000 keV. The observed yield clearly demonstrates the production of secondary particles from high-energy cosmic muons interacting with the Pb shielding. The statistical significance of this observation is 12.3 $\left( = \scriptstyle~\mathrm{N^{cor}_{MI}/\sqrt{N^{W}_{PD}(Corr.)}}\right)$.

\begin{table}[h]
  \small
  \centering
  \caption{\label{tab2}Systematic errors in the experimental data and Geant4 simulations, propagated to muon-induced secondary $\gamma$ and neutron events.\\}      
  \begin{tabular}{c|c|c|c}
    \hline
    {\bf Error} & \bf Set1 & \bf Set2 & \bf Systematic  \\ 
    {\bf source} &  &  & \bf error (\%)  \\
    \hline
    {$E_\mathrm{{dep}}$} &  30 < $E_\mathrm{{dep}}$ (keV) < 1450   &    30 < $E_\mathrm{{dep}}$ (keV) < 1350      & 1 \\
    \hline
    {$\Delta t_\mathrm{B-T}$} &  -8 < $\Delta t_\mathrm{B-T}$ (ns) < +8 & -4 < $\Delta t_\mathrm{B-T}$ (ns) < +4 & 1 \\
    \hline
    {$\Delta t_\mathrm{HPGe-PS}$} & 10 < $\Delta t_\mathrm{HPGe-PS}$ ($\mu$s) < 110  & 10 < $\Delta t_\mathrm{HPGe-PS}$ ($\mu$s) < 90  & 7.1  \\
    \hline
    {Efficiency (\%)}    & 47      & 43        & 4.5                         \\
    \hline
    {\bf{Total}}    &          &           &                          \\
    {\bf{systematic}}    & $-$         & $-$           & 8.5                         \\
    {\bf{error}}    &          &           &                          \\
    \hline
  \end{tabular}
\end{table}


Futhermore, the muon-induced secondary yield ($Y_\mathrm{MI}$) in Pb has been determined using the HPGe detector. This quantity represents the average number of secondaries generated per muon interaction with the top Pb shielding. The production rate can be estimated, with the information of the average distance ($L_\mathrm{avg}$) traversed by muons and the average interaction density ($\mathrm{\rho_{avg}}$) of muons through the Pb shielding. The parameter $\eta$ represents the ratio of detected muon-induced secondary events in the HPGe detector to the total number of secondary events generated by cosmic muons, and is estimated to be 52\% based on Geant4 MC simulations.

The average path length $L_\mathrm{avg}$ of cosmic muons in the Pb shielding is estimated from Geant4 MC simulation to be $\sim$12.5 cm. The density of the Pb used in the shielding is 11.3 g$\thinspace$cm$^{-3}$. 
Based on these values, the muon-induced secondary yield, $Y_\mathrm{MI}$, is given by:

\begin{equation}\label{event6}
  Y_\mathrm{MI} = \left( \frac{N^{\mathrm{cor}}_\mathrm{MI}}{\eta N_\mu L_{\mathrm{avg}} \rho_{\mathrm{avg}}} \right)
\end{equation}

where, $N^\mathrm{cor}_\mathrm{MI}$ represents the measured efficiency-corrected number of secondary events per day, as obtained in this study, arising from cosmic muon interactions with the composite top shielding. The quantity $N_\mathrm{\mu}$ corresponds to the number of cosmic muons passing through the top Pb shielding per day. Therefore, the yield of secondary events in the HPGe detector due to the Pb shielding is calculate to be $(11 \pm 1 (\text{ stat.}) \pm 1 (\text{ sys.}))$ secondary events$\thinspace\text{kg}^{-1}\thinspace\mathrm{m^{-2}}\thinspace\text{muon}^{-1}$.



\section{\label{sec6}Conclusions and Outlook}
In above-ground rare event search experiments, one of the most significant irreducible backgrounds arises from secondary particles produced by high-energy cosmic muons interacting with high-Z shielding materials, such as Pb. To investigate this, a HPGe detector was characterized using various radioactive sources, resulting a photopeak efficiency of 3\% at 274 keV and an energy resolution of 0.4\% at 662 keV. The HPGe detector was then employed, in coincidence with two plastic scintillator detectors, to measure the yield of cosmic muon-induced secondaries. A time-coincidence technique was utilized to reject these events, demonstrating the effectiveness of this approach in studying and quantifying muon-induced backgrounds in Pb shielding materials.

An excess of muon-induced secondaries has been observed in the experimental data in the presence of Pb shielding. The time responses of these secondary events have been compared with Geant4-based MC simulations. The measured mean characteristic time of $\Delta t_\mathrm{HPGe-PS}$ distribution is 11 $\pm$ 4 $\mathrm{\mu}$s, which is in good agreement with the MC simulation result of 11 $\pm$ 1 $\mathrm{\mu}$s. The event reconstruction efficiency of the detector is estimated to be 93$\%$ using Geant4 simulations. The measured rate of muon-induced secondaries is 34 $\pm$ 1 (stat.) $\pm$ 3 (sys.) events per day in the HPGe detector, with a statistical significance of 12.3. By applying a time-coincidence technique with a 100 $\mu$s window, the muon-induced secondary $\gamma$ and neutron backgrounds can be mitigated by 97$\%$, while introducing a dead time of less than 1$\%$ in the experimental setup. The corresponding yield of the muon-induced secondaries in Pb is determined to be $(11\pm 1 (\text{ stat.}) \pm 1 (\text{ sys.}))\thinspace\text{kg}^{-1}\thinspace\mathrm{m^{-2}}\thinspace\text{muon}^{-1}$, within the energy range of 30 keV to 2000 keV.

These results provide valuable inputs for designing background mitigation strategies in rare event searches, conducted with above-ground or shallow-overburden detector systems. Further suppression can be achieved by deploying an active veto system surrounding the HPGe detector to tag and reject those secondary events. Additionally, advanced machine learning algorithms~\cite{MLP} can be implemented to mitigate these backgrounds on an event-by-event basis, thereby reduce experimental uncertainties and enhance the physics sensitivity for detecting true CE$\nu$NS events in the HPGe detector.

\section{Acknowledgments}
We thank the Department of Atomic energy (DAE), India and the Department of Science and Technology (DST), India for financial support. We thank Mr. Lalatendu Mishra, Mr. Avneesh Kumar Tripathy, Mr. Debasis Barik and Mr. Deepak Kumar for their logistical support in setting up the experiment. The work is also funded through the J.C. Bose fellowship of DST awarded to Prof. Mohanty. We would also like to acknowledge the use of the Garuda and Kanaad HPC cluster facilities at SPS, NISER.

\bibliographystyle{JHEP}
\bibliography{references}

\providecommand{\href}[2]{#2}\begingroup\raggedright\begin{thebibliography}{10}

\bibitem{cevns}
D.Z.~Freedman, \emph{Coherent effects of a weak neutral current},
  \href{https://doi.org/10.1103/PhysRevD.9.1389}{\emph{Phys. Rev. D} {\bfseries
  9} (1974) 1389}.

\bibitem{coherent}
D.~Akimov, J.B.~Albert, P.~An, C.~Awe, P.S.~Barbeau, B.~Becker et~al.,
  \emph{Observation of coherent elastic neutrino-nucleus scattering},
  \href{https://doi.org/10.1126/science.aao0990}{\emph{Science} {\bfseries 357}
  (2017) 1123}
  [\href{https://arxiv.org/abs/https://www.science.org/doi/pdf/10.1126/science.aao0990}{{\ttfamily
  https://www.science.org/doi/pdf/10.1126/science.aao0990}}].

\bibitem{coherency}
{\scshape TEXONO Collaboration} collaboration, \emph{Coherency in
  neutrino-nucleus elastic scattering},
  \href{https://doi.org/10.1103/PhysRevD.93.113006}{\emph{Phys. Rev. D}
  {\bfseries 93} (2016) 113006}.

\bibitem{conus+}
H.~Bonet, A.~Bonhomme, C.~Buck, K.~F{\"u}lber, J.~Hakenm{\"u}ller, J.~Hempfling
  et~al., \emph{Full background decomposition of the conus experiment},
  \href{https://doi.org/10.1140/epjc/s10052-023-11240-4}{\emph{The European
  Physical Journal C} {\bfseries 83} (2023) 195}.

\bibitem{conus}
{\scshape CONUS Collaboration} collaboration, \emph{Final conus results on
  coherent elastic neutrino-nucleus scattering at the brokdorf reactor},
  \href{https://doi.org/10.1103/PhysRevLett.133.251802}{\emph{Phys. Rev. Lett.}
  {\bfseries 133} (2024) 251802}.

\bibitem{texono}
{\scshape TEXONO Collaboration} collaboration, \emph{New limits on the coherent
  neutrino-nucleus elastic scattering cross section at the kuo-sheng
  reactor-neutrino laboratory},
  \href{https://doi.org/10.1103/PhysRevLett.134.121802}{\emph{Phys. Rev. Lett.}
  {\bfseries 134} (2025) 121802}.

\bibitem{texono_bkg}
{\scshape TEXONO Collaboration} collaboration, \emph{Neutron background
  measurements with a hybrid neutron detector at the kuo-sheng reactor neutrino
  laboratory}, \href{https://doi.org/10.1103/PhysRevC.98.024602}{\emph{Phys.
  Rev. C} {\bfseries 98} (2018) 024602}.

\bibitem{mu_lead1}
L.~Reichhart, A.~Lindote, D.~Akimov, H.~Araújo, E.~Barnes, V.~Belov et~al.,
  \emph{Measurement and simulation of the muon-induced neutron yield in lead},
  \href{https://doi.org/https://doi.org/10.1016/j.astropartphys.2013.06.002}{\emph{Astroparticle
  Physics} {\bfseries 47} (2013) 67}.

\bibitem{mu_lead2}
R.~Dey, P.~Netrakanti, D.~Mishra, S.~Behera, R.~Sehgal, V.~Jha et~al.,
  \emph{Measurement of cosmic muon-induced neutron background with ismran
  detector in a non-reactor environment},
  \href{https://doi.org/https://doi.org/10.1016/j.astropartphys.2025.103101}{\emph{Astroparticle
  Physics} {\bfseries 169} (2025) 103101}.

\bibitem{wimp}
V.~P{\v{e}}{\v{c}}, V.A.~Kudryavtsev, H.M.~Ara{\'u}jo and T.J.~Sumner,
  \emph{Muon-induced background in a next-generation dark matter experiment
  based on liquid xenon},
  \href{https://doi.org/10.1140/epjc/s10052-024-12768-9}{\emph{The European
  Physical Journal C} {\bfseries 84} (2024) 481}.

\bibitem{dm_bkg}
K.~Bikit, D.~Mrdja, I.~Bikit and M.~Veskovic, \emph{Investigation of cosmic-ray
  muon induced processes by the miredo facility},
  \href{https://doi.org/https://doi.org/10.1016/j.apradiso.2013.11.102}{\emph{Applied
  Radiation and Isotopes} {\bfseries 87} (2014) 77}.

\bibitem{canberra}
online,
  \url{https://www.mirion.com/products/technologies/spectroscopy-scientific-analysis/gamma-spectroscopy/detectors/hpge-detectors-accessories/germanium-detectors}.

\bibitem{caen}
CAEN, ``V1730 digitizer.'' online, \url{https://www.caen.it/products/v1730/}.

\bibitem{compass}
CAEN, ``Compass.'' online, \url{https://www.caen.it/products/compass/}.

\bibitem{root}
R.~Brun and F.~Rademakers, \emph{Root — an object oriented data analysis
  framework},
  \href{https://doi.org/https://doi.org/10.1016/S0168-9002(97)00048-X}{\emph{Nuclear
  Instruments and Methods in Physics Research Section A: Accelerators,
  Spectrometers, Detectors and Associated Equipment} {\bfseries 389} (1997)
  81}.

\bibitem{geant4}
S.~Agostinelli, J.~Allison, K.~Amako, J.~Apostolakis, H.~Araujo, P.~Arce
  et~al., \emph{Geant4—a simulation toolkit},
  \href{https://doi.org/https://doi.org/10.1016/S0168-9002(03)01368-8}{\emph{Nuclear
  Instruments and Methods in Physics Research Section A: Accelerators,
  Spectrometers, Detectors and Associated Equipment} {\bfseries 506} (2003)
  250}.

\bibitem{cry}
``Cry.'' online, \url{http://nuclear.llnl.gov/simulation}.

\bibitem{brem_bkg}
H.~Bae, E.~Jeon, Y.~Kim and S.~Lee, \emph{Neutron and muon-induced background
  studies for the amore double-beta decay experiment},
  \href{https://doi.org/https://doi.org/10.1016/j.astropartphys.2019.06.006}{\emph{Astroparticle
  Physics} {\bfseries 114} (2020) 60}.

\bibitem{pho_bkg}
G.L.~Cassiday, \emph{Photonuclear interactions of high-energy muons},
  \href{https://doi.org/10.1103/PhysRevD.3.1109}{\emph{Phys. Rev. D} {\bfseries
  3} (1971) 1109}.

\bibitem{spa_bkg}
S.W.~Li and J.F.~Beacom, \emph{First calculation of cosmic-ray muon spallation
  backgrounds for mev astrophysical neutrino signals in super-kamiokande},
  \href{https://doi.org/10.1103/PhysRevC.89.045801}{\emph{Phys. Rev. C}
  {\bfseries 89} (2014) 045801}.

\bibitem{cap_bkg_1}
J.~Hadermann and K.~Junker, \emph{Emission of neutrons following muon capture
  in heavy nuclei},
  \href{https://doi.org/https://doi.org/10.1016/0375-9474(76)90250-5}{\emph{Nuclear
  Physics A} {\bfseries 271} (1976) 401}.

\bibitem{rec_bkg}
Z.~Chen, X.~Zhang, Z.~Yu, J.~Cao and C.~Yang, \emph{Radiogenic neutron
  background in reactor neutrino experiments},
  \href{https://doi.org/10.1103/PhysRevD.104.092006}{\emph{Phys. Rev. D}
  {\bfseries 104} (2021) 092006}.

\bibitem{Eff}
S.~Thakur, S.~Devi, S.S.~Kaintura, K.~Tiwari and P.P.~Singh,
  \emph{Spectroscopic performance evaluation and modeling of a low background
  hpge detector using geant4},
  \href{https://doi.org/https://doi.org/10.1016/j.nima.2023.168826}{\emph{Nuclear
  Instruments and Methods in Physics Research Section A: Accelerators,
  Spectrometers, Detectors and Associated Equipment} {\bfseries 1058} (2024)
  168826}.

\bibitem{reso}
R.~Cooper, M.~Amman, P.~Luke and K.~Vetter, \emph{A prototype high purity
  germanium detector for high resolution gamma-ray spectroscopy at high count
  rates},
  \href{https://doi.org/https://doi.org/10.1016/j.nima.2015.05.053}{\emph{Nuclear
  Instruments and Methods in Physics Research Section A: Accelerators,
  Spectrometers, Detectors and Associated Equipment} {\bfseries 795} (2015)
  167}.

\bibitem{comp}
N.~Dokania, V.~Singh, S.~Mathimalar, V.~Nanal, S.~Pal and R.~Pillay,
  \emph{Characterization and modeling of a low background hpge detector},
  \href{https://doi.org/https://doi.org/10.1016/j.nima.2014.01.064}{\emph{Nuclear
  Instruments and Methods in Physics Research Section A: Accelerators,
  Spectrometers, Detectors and Associated Equipment} {\bfseries 745} (2014)
  119}.

\bibitem{bkg_1}
E.~Andreotti, C.~Arnaboldi, F.~Avignone, M.~Balata, I.~Bandac, M.~Barucci
  et~al., \emph{Muon-induced backgrounds in the cuoricino experiment},
  \href{https://doi.org/https://doi.org/10.1016/j.astropartphys.2010.04.004}{\emph{Astroparticle
  Physics} {\bfseries 34} (2010) 18}.

\bibitem{bkg_2}
G.A.e.a.~Krishnamoorthy~H., Gupta~G., \emph{Study of $\gamma$-ray background
  from cosmic muon induced neutrons},
  \href{https://doi.org/https://doi.org/10.1140/epja/i2019-12822-3}{\emph{The
  European Physical Journal A} {\bfseries 55} (2019) }.

\bibitem{bkg_sub_1}
M.~Andriamirado, A.B.~Balantekin, H.R.~Band, C.D.~Bass, D.E.~Bergeron,
  N.S.~Bowden et~al., \emph{Prospect-ii physics opportunities},
  \href{https://doi.org/10.1088/1361-6471/ac48a4}{\emph{Journal of Physics G:
  Nuclear and Particle Physics} {\bfseries 49} (2022) 070501}.

\bibitem{bkg_sub_2}
P.~Netrakanti, D.~Mulmule, D.~Mishra, S.~Behera, R.~Dey, R.~Sehgal et~al.,
  \emph{Measurements using a prototype array of plastic scintillator bars for
  reactor based electron anti-neutrino detection},
  \href{https://doi.org/https://doi.org/10.1016/j.nima.2021.166126}{\emph{Nuclear
  Instruments and Methods in Physics Research Section A: Accelerators,
  Spectrometers, Detectors and Associated Equipment} {\bfseries 1024} (2022)
  166126}.

\bibitem{MLP}
D.~Mulmule, P.~Netrakanti, L.~Pant and B.~Nayak, \emph{Machine learning
  technique to improve anti-neutrino detection efficiency for the ismran
  experiment},
  \href{https://doi.org/10.1088/1748-0221/15/04/P04021}{\emph{Journal of
  Instrumentation} {\bfseries 15} (2020) P04021}.

\end{thebibliography}\endgroup

\end{document}